\documentclass{aa}
\usepackage{graphicx}
\usepackage{natbib}
\usepackage{txfonts}
\usepackage{amssymb}
\bibpunct{(}{)}{;}{a}{}{,} 
\begin{document}

   \title{Horizontal supergranule-scale motions inferred from TRACE ultraviolet observations of the chromosphere}
   \author{H. Tian\inst{1}
           \and
           H. E. Potts\inst{2}
           \and
           E. Marsch\inst{3}
           \and
           R. Attie\inst{3}
           \and
           J.-S. He\inst{3}
          }
   \offprints{H. Tian}
   \institute{School of Earth and Space Sciences, Peking University, Beijing, China\\
             \email{tianhui924@pku.edu.cn}
             \and
             Department of Physics and Astronomy, University of Glasgow, G12 8QQ, Glasgow, UK\\
             \email{hugh@astro.gla.ac.uk}
             \and
             Max-Planck-Institut f\"ur Sonnensystemforschung, Katlenburg-Lindau, Germany\\
              \email{marsch@mps.mpg.de}
             }
   \date{}

\abstract
{}
{We study horizontal supergranule-scale motions revealed by TRACE
observation of the chromospheric emission, and investigate the
coupling between the chromosphere and the underlying photosphere.}
{A highly efficient feature-tracking technique called balltracking
has been applied for the first time to the image sequences obtained
by TRACE (Transition Region and Coronal Explorer) in the passband of
white light and the three ultraviolet passbands centered at
1700~\AA, 1600~\AA, and 1550~\AA. The resulting velocity fields have
been spatially smoothed and temporally averaged in order to reveal
horizontal supergranule-scale motions that may exist at the emission
heights of these passbands.}
{We find indeed a high correlation between the horizontal velocities
derived in the white-light and ultraviolet passbands. The horizontal
velocities derived from the chromospheric and photospheric emission
are comparable in magnitude.}
{The horizontal motions derived in the UV passbands might indicate
the existence of a supergranule-scale magneto-convection in the
chromosphere, which may shed new light on the study of mass and
energy supply to the corona and solar wind at the height of the
chromosphere. However, it is also possible that the apparent motions
reflect the chromospheric brightness evolution as produced by
acoustic shocks which might be modulated by the photospheric
granular motions in their excitation process, or advected partly by
the supergranule-scale flow towards the network while propagating
upward from the photosphere. To reach a firm conclusion, it is
necessary to investigate the role of granular motions in the
excitation of shocks through numerical modeling, and future
high-cadence chromospheric magnetograms must be scrutinized.}


\titlerunning{Horizontal motions inferred from TRACE observations}
\authorrunning{H. Tian et al.}
\maketitle

\section{Introduction}

Quasi-steady convective flows of different scales have been
suggested to play an important role in the processes of mass supply
and energy transport across different layers of the solar atmosphere
\citep[e.g.,][]{Foukal1978,Krijger2002,Marsch2008,Dammasch2008,Curdt2008}.
As is well known, granulation with a typical 1-Mm size and 5-minute
lifetime are ubiquitous in high-resolution photospheric images
\citep[e.g.,][]{Title1989,Berrilli2002,Jin2009}. Groups of theses
granules tend to move in a systematic way, which is characterized by
a cellular convective motion at scales of the order of 32~Mm
\citep{Leighton1962}. This larger-scale flow was termed
supergranulation, with cells of which the boundaries coincide with
the chromospheric network and lanes of magnetic concentrations
\citep[e.g.,][]{Leighton1962,Simon1964,Simon1988,WangZirin1988}.

Under the assumption that the granules can be considered as tracers
of the underlying larger-scale velocity fields
\citep{Simon1988,Rieutord2001} of magnetoconvection, one can derive
the related horizontal flow speed by the widely used LCT (Local
correlation tracking) technique \citep{November1988}. Recently,
\cite{Potts2004} developed a new method for tracking flow fields.
This so called balltracking method is more noise-tolerant and highly
efficient for analysing large data sets. The flow field inferred by
this method has a similar accuracy to that obtained by LCT.

The horizontal motions in the chromosphere have not been scrutinized
and are not understood well. Few attempts have been made to measure
chromospheric proper motions, e.g., by applying LCT to the H$\alpha$
images obtained in active regions \citep{Yi1995,Chae2000,Yang2003}.
By using this method, the typical horizontal velocity in the
chromospheric network was found to be 1000-1500~m/s \citep{Yi1995}.
However, the small-scale structures present in H$\alpha$ images
appear to be elongated and thus are different from the
cellular-shaped granules. Moreover, the H$\alpha$ line is very
sensitive to and strongly influenced by dynamic events. Thus, the
explanation of the horizontal velocities as measured from H$\alpha$
images is not straightforward.

Besides H$\alpha$, the Ca~{\sc{ii}} H \& K lines have also been
extensively used to explore dynamics in the chromosphere. The
chromospheric network, which coincides with the magnetic network and
outlines the supergranular boundaries, is the most prominent
structure on images of Ca~{\sc{ii}}, while the cell interiors
(internetwork regions) which are enclosed by the network occupy most
of the quiet-Sun area. Internetwork areas are filled with
intermittent grainy brightenings. These emission features are
usually termed H(K)$_{2v}$ cell grains because their intensities
peak dramatically just blueward of the Ca~{\sc{ii}} H \& K line
centers. These internetwork grains (or cell grains) are also present
on wider-band (a few {\AA}) Ca~{\sc{ii}} filtergrams, but at a lower
contrast and a slight phase shift \citep{Rutten1999a,Rutten1999b}.
The sizes of internetwork grains are about 1$\sim$4~Mm, and their
lifetimes are usually less than 12 minutes
\citep{Rutten1991,Tritschler2007}. For a detailed review of the
Ca~{\sc{ii}} H$_{2v}$ \& K$_{2v}$ cell grains, we refer to
\cite{Rutten1991}.

By comparing the quiet-Sun images obtained by TRACE
\citep[Transition Region and Coronal Explorer,][]{Handy1999} in its
three ultraviolet (UV) passbands (centered at 1700~\AA, 1600~\AA,
and 1550~\AA) and a co-temporal Ca~{\sc{ii}} K filtergram,
\cite{Rutten1999a} found a high degree of spatial agreement between
the Ca~{\sc{ii}} K image and the TRACE images, in particular the
1700~\AA~one. They concluded that the three TRACE channels portray
internetwork grain phenomena as well as Ca~{\sc{ii}} K filtergrams
do. Therefore, TRACE image sequences provide an excellent means to
study the dynamics of internetwork grains \citep{Rutten1999a}.

By applying here the balltracking technique for the first time to
the TRACE images obtained in the passband of white light and the
three UV passbands centered at 1700~\AA, 1600~\AA, and 1550~\AA, we
have made an attempt to study horizontal photospheric and
chromospheric motions in a large area of the quiet Sun. Our analysis
revealed a striking correlation between the horizontal velocities
derived in the white-light passband and the UV passbands, a finding
which might shed new light on our understanding of mass supply to
the corona and solar wind. Alternatively, our finding may provide
insights into the excitation and propagation of the shock sequences
which might produce the bright internetwork grains.

\section{Data analysis and results}

The TRACE image sequence analysed here was obtained on 14 October 1998
from 08:20 to 09:30 UTC in a quiet-Sun region around disk center. Images
in the four passbands as mentioned above were taken with a cadence of 21~s.
The sizes of the images are 512$\times$512~pixels squared, with a pixel size
of 0.5$^{\prime\prime}$.

\begin{figure*}
\centering {\includegraphics[width=\textwidth]{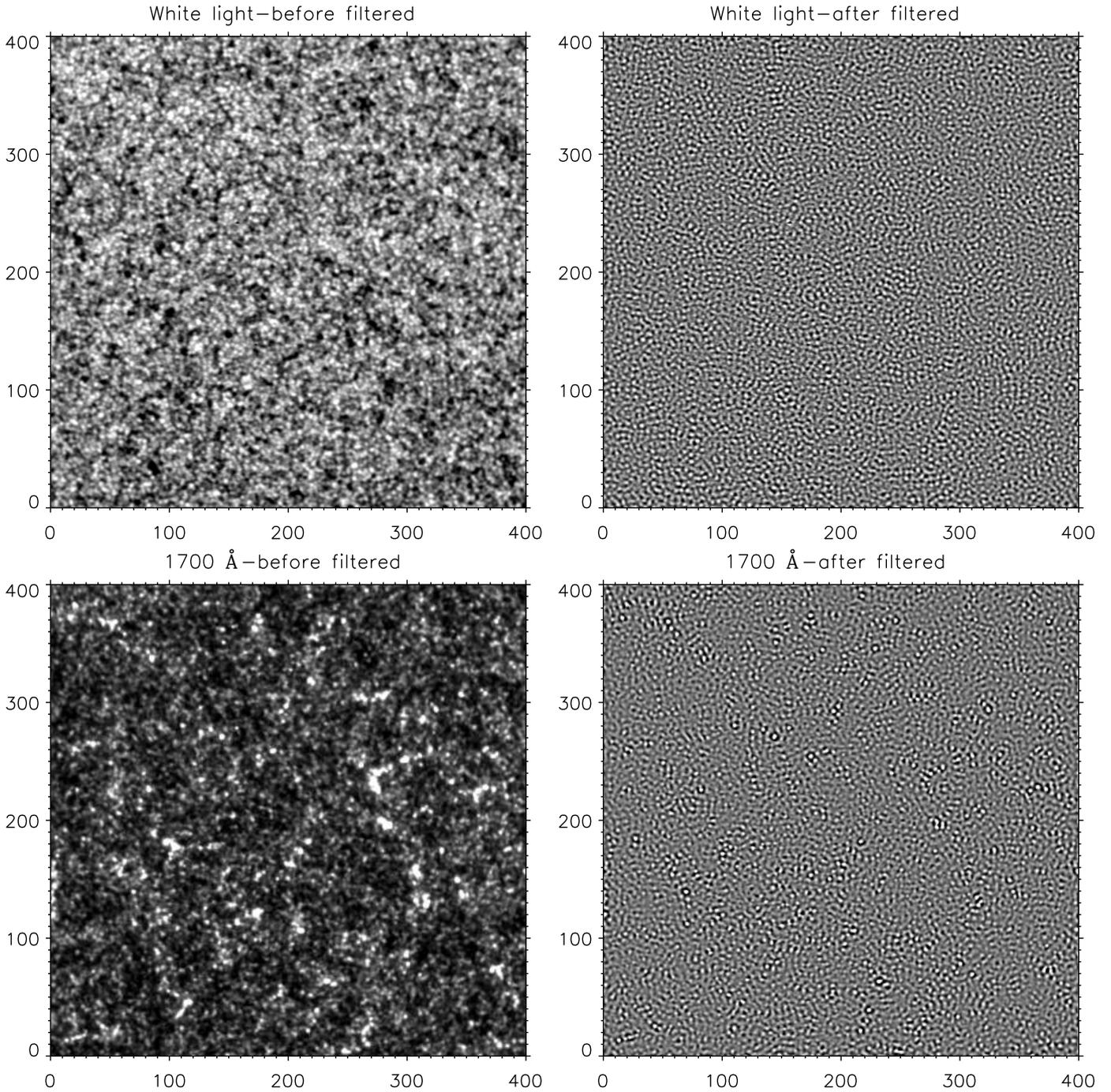}}
\caption{Images taken at 08:20 in the passbands of white light
(upper) and 1700~\AA~(lower). The original and filtered images are
presented on the left and right, respectively. The size of each
image is $200^{\prime\prime}\times200^{\prime\prime}$. }
\label{fig.1}
\end{figure*}

The standard software for reducing TRACE data was applied to this
data set, including the removal of cosmic rays, subtraction of dark
currents, normalization of the counts, and so on. Then we used the
IDL procedure \textit{Rot\_XY.pro} available in \textit{SolarSoft}
to take into account the solar rotation effect. The satellite jitter
was subsequently removed with an accuracy of $0.025^{\prime\prime}$,
by applying the cross-correlation technique to enlarged (by a factor
of 20) images. For computational efficiency, we resampled the data
to obtain a cadence of 42~s, and then extracted a sub-region with a
size of $200^{\prime\prime}\times200^{\prime\prime}$ from the
coaligned images.

Each image was spatially Fourier filtered to remove solar
oscillations and observational noise and thus to remove the large
scale variations, as described in \cite{Potts2004}. In addition, the
1700~\AA, 1600~\AA, and 1550~\AA~images have the problem that the
intensity variation is very non-linear, with a small number of very
bright points that tend to dominate the filtered data. These data
were scaled by raising the intensity to a power of 0.4 to compress
the intensity data into a more uniform range, before filtering. We
kept the Fourier components with wavelength between
$1^{\prime\prime}$ (the minimum measurable wavelength) and
$3.5^{\prime\prime}$ (upper limit of the granule/grain size).
Figure~\ref{fig.1} presents the first images of the sequences taken
with the passbands of white light and 1700~\AA. It is clear from the
figure that the size of the chromospheric grains inside cell
interiors is comparable to that of the photospheric grains. The
brightness range of the cell grains seems to be larger than that of
the photospheric grains. The most prominent difference between the
photospheric emission and the chromospheric emission is the much
more enhanced network emission in the chromosphere. But the network
bright points only constitute a minor part of the area, and thus
their contribution to the tracked velocities is minor.
Figure~\ref{fig.1} also shows the corresponding filtered images,
which reveal clearly that both the dark and bright points inside the
cell interiors are comparable in size (or width).

In the balltracking technique as developed by \cite{Potts2004}, one
considers a photospheric image in a three-dimensional representation
by regarding the brightness as corresponding to a geometrical
height. Small floating balls are dropped on to this surface and tend
to settle in the local minima between some adjacent granules. As the
granulation pattern evolves, the balls will be pushed around and
thus follow and reveal the motion of the granules. It has been
demonstrated that the results from this method have an accuracy
similar to those produced by LCT \citep{Potts2004}. This method has
been successfully applied to high-resolution continuum images
obtained by MDI/SOHO \citep{Potts2007,Potts2008,Innes2009} and
SOT/Hinode \citep{Attie2009}.

As mentioned by \cite{Potts2004}, the balltracking method is not
limited to flows in the photosphere, but will also work for any
velocity field in which there are visible moving features of known
scale length. Here we made the first attempt to apply this method to
TRACE images in the passband of white light and several UV
passbands. We set the radius of the tracking ball at
$1.2^{\prime\prime}$. As described in \cite{Attie2009}, the output
of the balltracking procedure are velocity fields derived from the
displacements of many individual tracking balls, which reflect the
fast and stochastic small-scale granular motions. To extract the
underlying large-scale velocity field the data needs to be spatially
smoothed and temporally averaged. \cite{Rieutord2001} suggested that
the velocity field may not be faithfully described by granular
motions at spatial scales less than 2.5~Mm, or at temporal scales
shorter than half an hour. Our tracking results have been smoothed
over 3.5~Mm and running-averaged over the entire 70 minutes. The
tracking routine actually tracked the data twice, once with the
normal data and once with the inverted data, and gave very similar
results in both cases. We took the average of the two results. The
error estimate for the tracking was obtained from the difference of
these two results. The root mean square values of the errors are
well below those of the tracked velocities in all of the four
passbands.

\begin{figure}
\centering
{\includegraphics[height=0.3\textheight,width=0.5\textwidth]{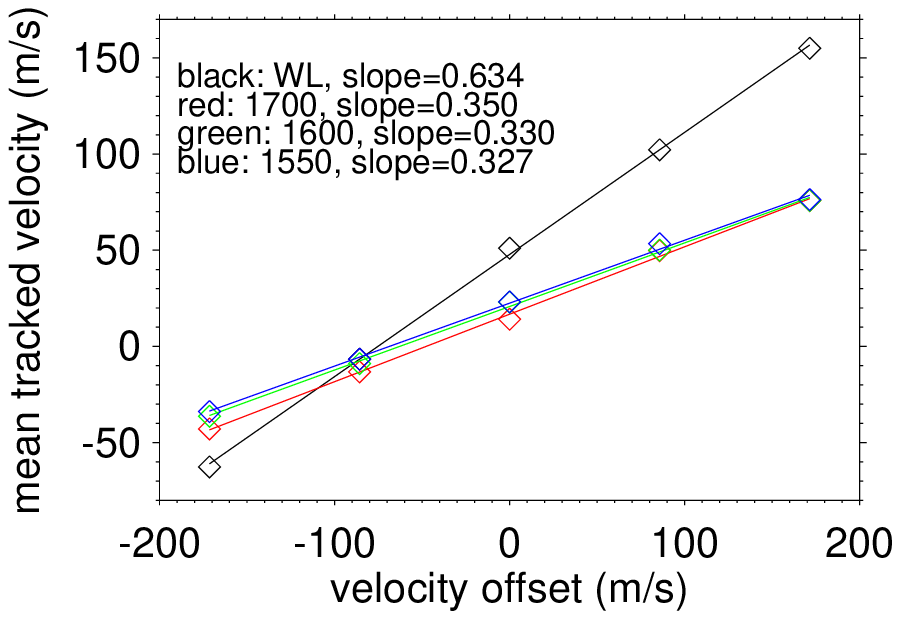}}
\caption{Calibration factors obtained by a linear fit to the mean
velocity derived from each offset data cube versus the offset
velocity, in the passbands of white light (black), 1700~\AA~(red),
1600~\AA~(green), and 1550~\AA~(blue). The values of the slopes are
used as velocity calibration factors. } \label{fig.2}
\end{figure}

Finally, since the velocity values obtained by balltracking are
usually underestimated as compared to the true velocities
\citep{Potts2004}, we need to divide the balltracked velocities by a
scaling factor. Starting with the de-rotated and jitter-removed data
cube, we chose a sub-area from the first frame of the data, and
offset each subsequent frame in turn by a small distance,
representing an additional imposed velocity. Four velocities were
chosen to produce four offset data cubes, which were then tracked
using balltracking. By applying a linear fit to the mean velocity
derived from each offset data cube versus the offset velocity, we
obtained a scaling factor in each passband. The derived velocities
were divided by this value. Figure~\ref{fig.2} shows the result of
the fitting in each passband. The values of the slopes are used as
velocity calibration factors.

\begin{figure}
\centering
{\includegraphics[height=0.3\textheight,width=0.5\textwidth]{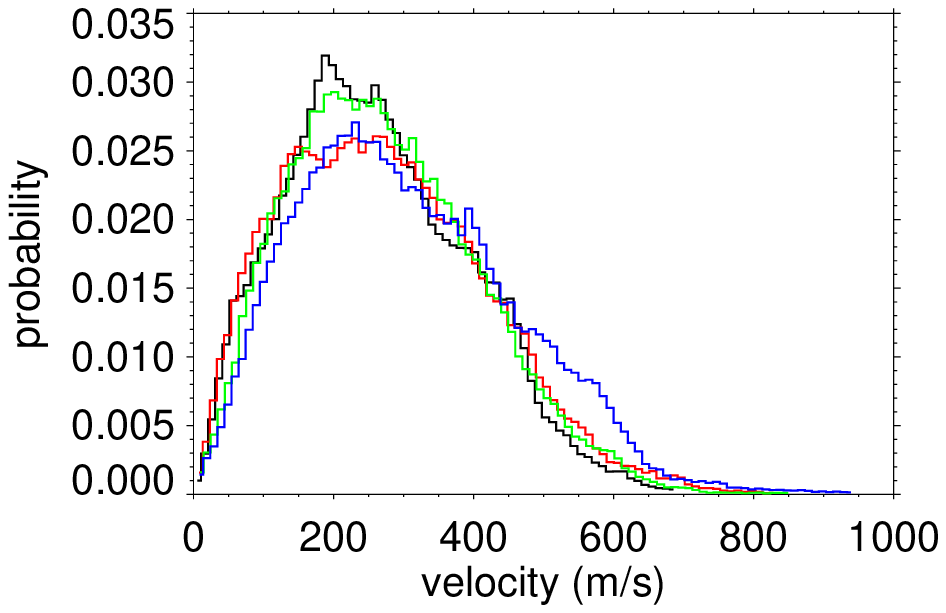}}
\caption{Histogram of the measured horizontal velocities in the
passbands of white light (black), 1700~\AA~(red), 1600~\AA~(green),
and 1550~\AA~(blue).} \label{fig.3}
\end{figure}

The distributions of the resulting velocities are presented in
Fig.~\ref{fig.3}. It turns out that the velocity values are
comparable in different passbands. The median values of the derived
velocities are 247~m/s, 258~m/s, 258~m/s, and 284~m/s, in the
passbands of white light, 1700~\AA, 1600~\AA, and 1550~\AA,
respectively. Following the method in \cite{Potts2004}, we estimated
an uncertainty of 29~m/s for the velocity measurement.

\begin{figure*}
\centering {\includegraphics[width=\textwidth]{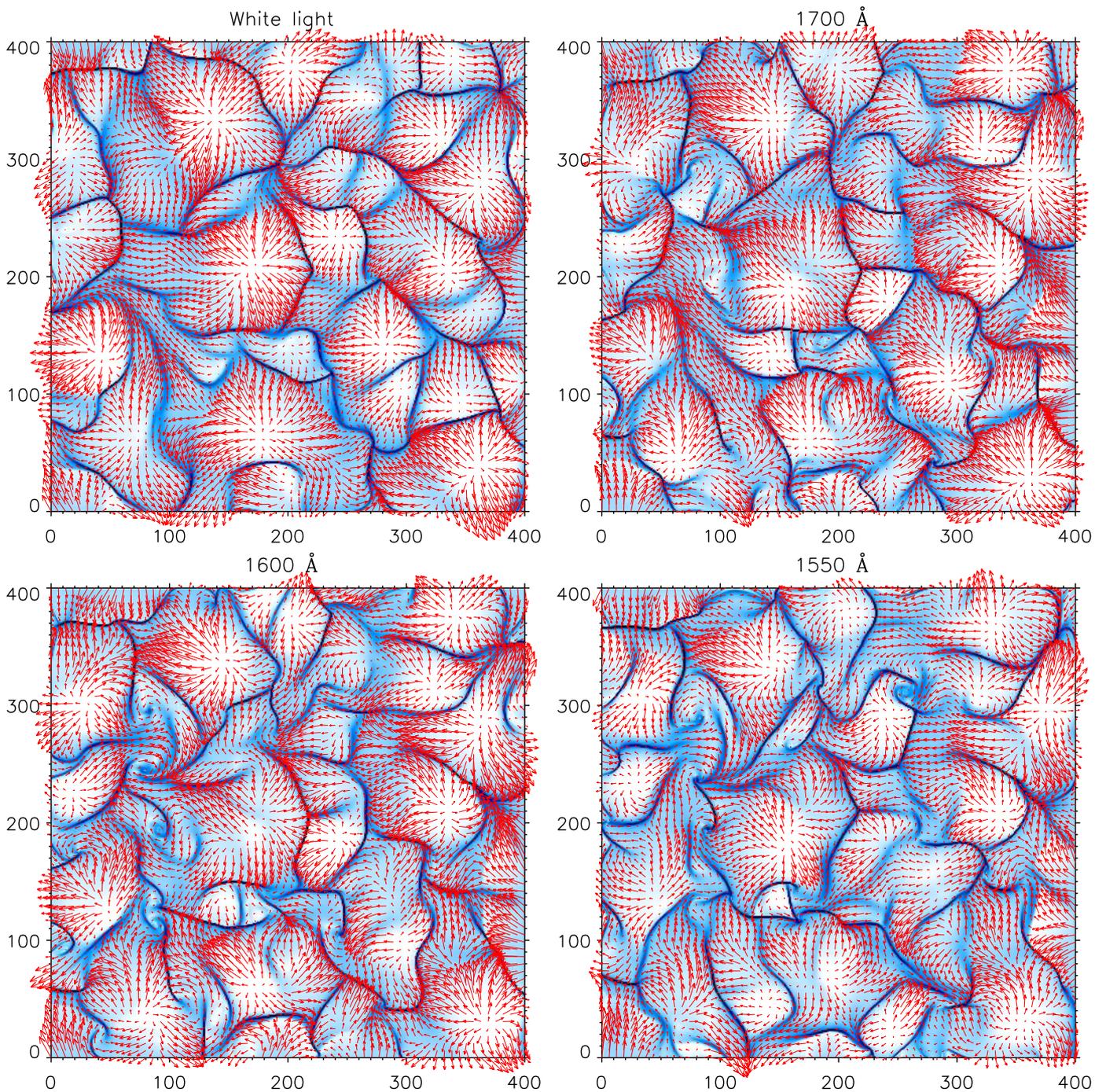}} \caption{
Velocity vector fields as calculated by using the image sequence in
the passband of white light (upper left), 1700~\AA~ (upper right),
1600~\AA~(lower left), and 1550~\AA~(lower right). The lanes
represent cell boundaries as calculated from the velocity field in
the corresponding passband. The size of each image is
$200^{\prime\prime}\times200^{\prime\prime}$.} \label{fig.4}
\end{figure*}

Figure~\ref{fig.4} presents the velocity fields for the four
passbands. The cell boundaries, which were calculated by applying
the automatic recognition algorithm developed by \cite{Potts2008} to
velocities obtained in the corresponding passband, are shown as the
background (lanes). A direct comparison of the cell boundaries as
calculated from the velocity field in white light and from that in
the passband of 1700~\AA~with the 1700~\AA~intensity image was made
and is shown in Fig.~\ref{fig.5}.

We also calculated the linear Pearson correlation coefficients of
each component of the velocity field as well as the velocity
magnitude, between each pair of passbands. The results are listed in
Table~\ref{table1}. The average correlation coefficients of the
unfiltered intensity images between each pair of passbands are also
listed therein.

\begin{table}[]
\caption[]{ The linear Pearson correlation coefficients of the
intensity (\textit{I}), two components of the velocity field
(\textit{$v_x$} and \textit{$v_y$}), as well as the velocity
magnitude (\textit{v}) between each pair of passbands.}
\label{table1}
\begin{center}
\begin{tabular}{p{2cm} p{1.0cm} p{1.0cm} p{1.0cm} p{1.0cm}}
\hline\hline
passbands  & \textit{I} &  \textit{$v_x$} & \textit{$v_y$} & \textit{v} \\
\hline
WL/1700   & 0.13 &  0.63 & 0.71 & 0.35 \\
WL/1600   & 0.09 &  0.58 & 0.62 & 0.26 \\
WL/1550   & 0.06 &  0.50 & 0.61 & 0.22 \\
1700/1600 & 0.96 &  0.81 & 0.81 & 0.63 \\
1700/1550 & 0.82 &  0.73 & 0.71 & 0.48 \\
1550/1600 & 0.92 &  0.78 & 0.81 & 0.61 \\
\hline
\end{tabular}
\end{center}
\end{table}

\section{Supergranular horizontal flows in the photosphere}

Photospheric horizontal flows have been extensively investigated
through ground based observations. Image sequences obtained by the
TRACE satellite are suitable for motion tracking in the photosphere
as they offer long uninterrupted runs over large areas (up to
$512^{\prime\prime}\times512^{\prime\prime}$) with no seeing
problems. The spatial resolution of $1^{\prime\prime}$ is just high
enough to resolve the granular structure.

\begin{figure*}
\centering {\includegraphics[width=\textwidth]{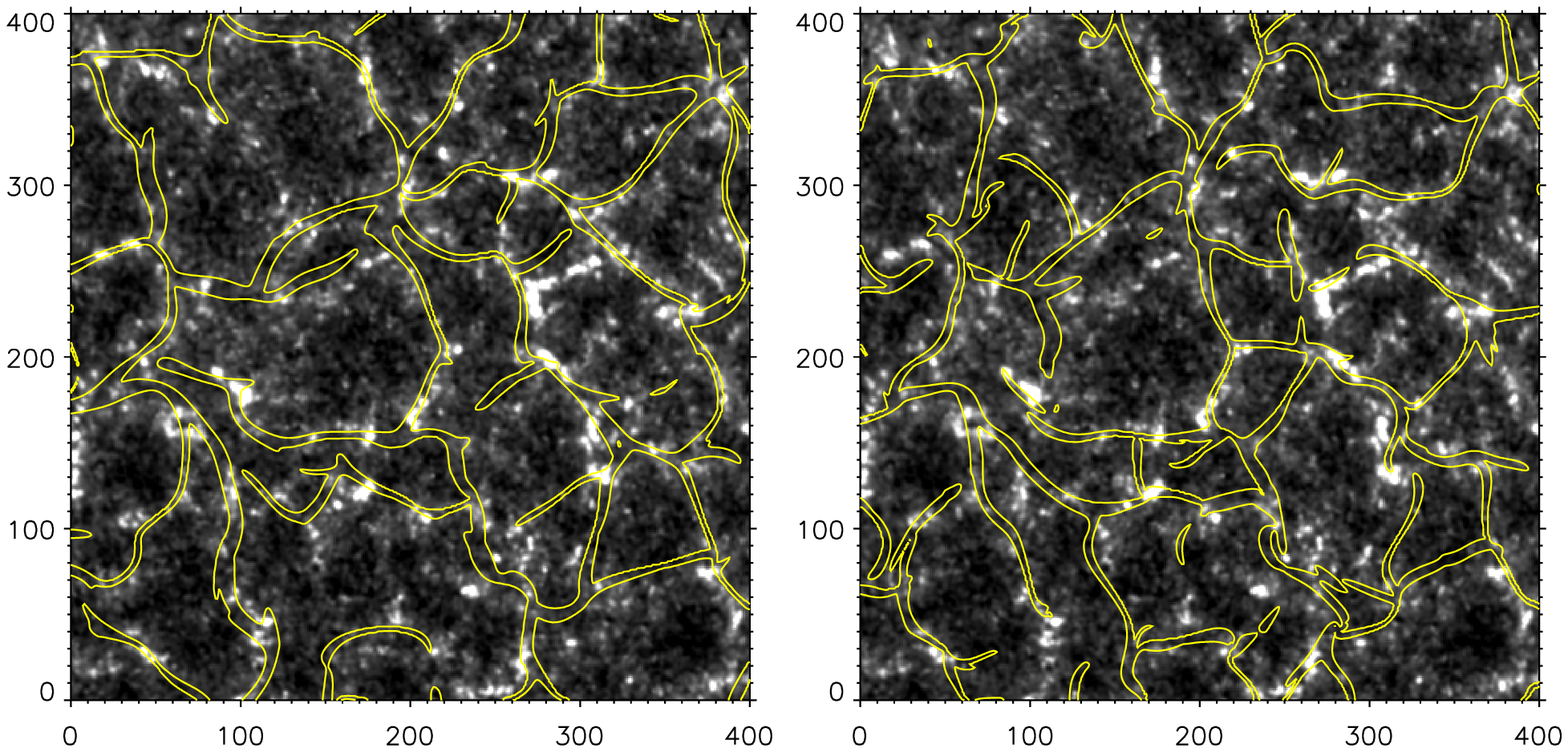}} \caption{The
yellow lanes representing cell boundaries as calculated from the
velocity field in white light (left) and from that in the passband
of 1700~\AA~(right) are superposed on the averaged intensity image
in the 1700~\AA~passband. The size of the image is
$200^{\prime\prime}\times200^{\prime\prime}$.} \label{fig.5}
\end{figure*}

By applying the balltracking technique to white light images
observed by TRACE, we have reconstructed the well-known photospheric
supergranular flow pattern in a quiet-Sun region with a size of
$200^{\prime\prime}\times200^{\prime\prime}$. From Fig.~\ref{fig.5}
we find that the calculated cell boundaries generally match the
network pattern as seen in the 1700~\AA~image. In the
1700~\AA~image, there are still a few segments of the bright lanes
which are not reproduced from these calculations. It might be due to
the effect of temporal variation, height variation, UV contamination
of the white-light emission, smoothing and averaging of the tracking
results, or the fact that small granules below the detection limit
of TRACE can not be traced properly.

The velocity magnitude in white light is consistent with the results
in \cite{Krijger2002} and \cite{Attie2009}, but smaller than those
in \cite{Wang1995} and \cite{Roudier1999}. This should be the result
of the different smoothing and averaging of the tracked velocity
fields. As mentioned in \cite{Attie2009}, a weaker smoothing and
averaging will increase the velocity values. Our choice of the
smoothing and averaging scales is a compromise between a more
accurate velocity measurement and a higher resolution to distinguish
flows with sharp gradient.

\cite{Krijger2002} applied the LCT technique to a TRACE white-light
image sequence in a small region, and concluded that it is possible
to use the TRACE observations to measure photospheric horizontal
velocity fields. Here we have demonstrated that the supergranular
flow pattern in the photosphere can be recovered by applying the
highly efficient balltracking method to the white-light images
observed by TRACE in a large area of the quiet Sun.

\section{Horizontal motions revealed through the chromospheric emission}

As mentioned previously, the horizontal flow field in the
chromosphere has not been understood well. Only a few attempts have
been made to measure chromospheric proper motions by applying LCT to
H$\alpha$ images obtained in active regions
\citep{Yi1995,Chae2000,Yang2003}. The TRACE observations allow a
direct comparison of possible large-scale horizontal motions in the
chromosphere with those in the underlying photosphere.

From Fig.~\ref{fig.4} we can see that the velocity fields obtained
in the three UV passbands also show cell structures which are
similar to those in the white-light passband. The diverging
flow-like patterns in some cells are highly coincident in different
passbands. The correlation coefficients listed in Table~\ref{table1}
are large, again indicating a high correlation between the derived
horizontal motions in different passbands. The cell boundaries
calculated from the velocity field in the 1700~\AA~passband also
coincide more or less with the chromospheric network, as shown in
Fig.~\ref{fig.5}. According to our knowledge, this is the first time
that supergranulation-scale horizontal motions, both in and above
the photosphere, are revealed in a large area of the quiet Sun.

\subsection{Formation height of emission from TRACE UV filters}

According to \cite{Handy1999} and \cite{Worden1999}, the emission in
the 1700~\AA~passband (4-10$\times$10$^3$~K) is UV continuum
originating around the temperature minimum; the emission in the
1600~\AA~passband (4-10$\times$10$^3$~K) has contributions from the
UV continuum and several lines like Fe~{\sc{ii}} and C~{\sc{i}}, but
the dominant contributor is the UV continuum from the region of
temperature minimum. According to the VAL-C solar atmosphere model
\citep{Vernazza1981}, the temperature minimum region is located at
about 500 km above $\tau_{500}$=1, corresponding to the upper
photosphere/lower chromosphere
\citep{Worden1999,Rutten2003,McAteer2004}.

The emission from the 1550~\AA~passband is mainly a combination of
C~{\sc{iv}} and the underlying UV continuum emission
\citep{Handy1999,Worden1999}. As mentioned in \cite{Krijger2001} and
\cite{Rutten1999a}, the C~{\sc{iv}} lines dominate the emission of
the 1550~\AA~passband in active regions, while in the quiet Sun
contributions from the UV continuum and several weak lines (e.g.
C~{\sc{i}} lines) dominate in the 1550~\AA~passband.
\cite{Handy1999} mentioned that the 1550~\AA~image resembles an
image taken in the center of the Ca~{\sc{ii}} K line with a bandpass
of a few Angstroms. So the major contributor to the emission of the
1550~\AA~passband seems to be the UV continuum, and thus the
1550~\AA~passband should be dominated by the chromospheric emission.

Similar to many papers such as \cite{Krijger2001} and
\cite{McAteer2004}, we use the term "chromosphere" even for the
upper photosphere since images of the TRACE UV passbands largely
differ from the photospheric images and essentially sample the
chromospheric network. We realize that there are limitations to
assigning the height of formation to emission from the broad
band-pass filters on the TRACE UV channels. However, by comparing
the quiet-Sun images obtained by TRACE in three UV passbands
(centered at 1700~\AA, 1600~\AA, and 1550~\AA) and a co-temporal
Ca~{\sc{ii}} K filtergram, \cite{Rutten1999a} concluded that there
is a high degree of spatial correspondence between the Ca~{\sc{ii}}
K image and the TRACE images, and that the three TRACE channels
reveal internetwork emission features as well as Ca~{\sc{ii}} K
filtergrams do. Similar conclusion was also reached by
\cite{Handy1999}. Thus, we believe that the structures visible on
the TRACE UV filters are well formed at a certain height, probably
the lower part, of the chromosphere.

The cross-talk (overlap) between emission from different passbands
might also influence our results. First, since our data were
obtained shortly after TRACE was launched, the transmission curves
of UV passbands presented in Fig. 9 of \cite{Handy1999} should have
not changed too much, and thus the leaking of the white-light signal
into the UV emission should be marginal. Second, there is a large
part of overlap between emission of the 1700~{\AA} and 1600~{\AA}
passbands, which accounts for the very large correlation between
velocities in the two passbands. This overlap does not change our
conclusion, since our main finding is the correlation between
velocities in the white-light and UV passbands, not the correlation
between two UV passbands. Third, the UV contamination of the
white-light emission might not be trivial and is likely to be one of
the reasons why some bright chromospheric network lanes are not
reconstructed by the calculated cell boundaries. This is, however,
counteracted by a higher correlation between the velocities than the
intensities.

\subsection{Nature of internetwork grains}

Internetwork grains (or cell grains) have been extensively studied
through ground based observations using Ca~{\sc{ii}} H \& K lines
\citep[for a review, see][]{Rutten1991}. As mentioned above,
\cite{Rutten1999a} found a high degree of spatial correspondence
between the Ca~{\sc{ii}} K image and the TRACE UV images (passbands
of 1700~\AA, 1600~\AA, and 1550~\AA). They concluded that the three
TRACE channels portray internetwork grain phenomena as well as
Ca~{\sc{ii}} K filtergrams do.

Internetwork regions occupy most of the quiet Sun area. So with the
balltracking technique we are mainly tracking the motions of the
internetwork grains in TRACE UV filters. Before interpreting the
observed horizontal motions in the chromosphere, we need to
understand the nature of these internetwork grains.

In fact, there are long-lasting disputes on this issue. Direct
evidence for the spatial correspondence between the cell grains and
internetwork magnetic elements \citep{Livingston1975} was provided
by \cite{Sivaraman1982}. The magnetic origin of the cell grains was
further confirmed by \cite{Dame1984}, \cite{Dame1985},
\cite{Dame1987}, \cite{Dame1988}, \cite{Sivaraman1991}, and
\cite{Sivaraman2000}. Based on \textit{Hinode} observations,
\cite{deWijn2008} found that some of the magnetic elements are
associated with the chromospheric cell grains. Recently, although no
obvious correlation between the magnetic flux density and the
Ca~{\sc{ii}} H internetwork brightness was obtained, \cite{Yang2009}
\textbf{were} still inclined to believe that there should be a
correlation between the two, and they attributed the lack of
correlation to the low cadence of Ca~{\sc{ii}} H images they used.

However, others doubt this conclusion. And through observations and
modelings they believe that the occurrence of internetwork grains
does not depend on magnetism but is caused by weak acoustic shocks
which propagate upwards through the low chromosphere
\citep{Rutten1991,Remling1996,Carlsson1997,Lites1999,Loukitcheva2009}.
\cite{Worden1999} compared the cell grains observed in TRACE UV
passbands and the corresponding photospheric magnetic field, and
suggested that the internetwork magnetic field is essentially
uninvolved with the production of cell grains in the 1600~\AA~
passband, but could have a small involvement with the cell grains in
the 1216~\AA~(Ly$\alpha$) passband.

\cite{Brandt1992,Brandt1994} found that there are two types of cell
grains: (i) oscillatory K$_{2v}$ flashers, appear only a few times
with a modulation of 3-minute oscillation, brief spatial memory,
5-10 grains per cell; (ii) persistent K-line grains, retaining
long-term identity while traveling through the cell, flashing with
3-5 minutes periodicity, about 1 grain per cell. The short-living
grains were suggested to be generated by acoustic shocks and have
nothing to do with the magnetic field. On the contrary, a spatial
correspondence between internetwork magnetic features and persistent
flashers was observed. The persistent flashers probably represent
the chromospheric signatures of newly-emerged strong-field ephemeral
regions on their ways to be part of the network. This finding was
confirmed by \cite{Nindos1998}. \cite{Lites1999} also found
persistent flashers but pointed out that they are very rare. While
with UV image sequences obtained by TRACE, \cite{Krijger2001}
identified many such long-living features, and \cite{McAteer2004}
also found that $\sim$50\% cell grains are persistent flashers.
Persistent flashers were called internetwork bright points by
\cite{deWijn2005}, in which they were suggested to correspond to the
strongest internetwork magnetic elements.

The question of whether or not the chromospheric cell grains are
correlated with the magnetic flux in internetwork regions is very
important to study the chromospheric heating and dynamics. Here we
don't intend to answer whether or not the magnetic field plays a
role, or a major role, in the production of chromospheric cell
grains, since there are still debates on this issue and it is not so
easy to distinguish between the two possibilities. As pointed out by
\cite{Kamio2006}, we need more detailed observations to solve this
problem. Instead, we discuss the implications of our results and
point out directions of future studies, with respect to each of the
two suggested natures of cell grains.

\subsection{Magneto-convection in the chromosphere}

As shown in Fig.~\ref{fig.4}, the measured horizontal velocity
fields in the chromosphere and photosphere show a similar
supergranule-scale cell-like pattern, which prompts us to think if
they represent an upward extension of supergranular horizontal flows
to the chromosphere.

The idea that supergranular flows may penetrate into higher layers
has already been suggested by several authors. It is known that
motions of granules in the photosphere follow the supergranulation
plasma flows because these plasma flows are observed in Doppler
shift. In the middle chromosphere (formation height of Si~{\sc{ii}}
$\lambda$1817), such Doppler shift was also investigated and
demonstrated to be existing \citep{November1979}.
\cite{November1979} even predicted that supergranular velocities
should be evident in the TR. \cite{November1982} found that
large-scale patterns of up- and downflows in the middle chromosphere
correlate well with those seen in the photosphere and concluded that
vertical flows of supergranular scale appear to extend into the
chromosphere. Height variation of the velocity field in the
photosphere has also been studied. Observations by
\cite{Deubner1971} indicated that the vertical motions increase with
height, and that the horizontal supergranular flow decreases
slightly, in the photosphere from the formation height of C~{\sc{i}}
$\lambda$5380 to that of Mg~{\sc{i}} $\lambda$5173. The latter is
formed at the temperature minimum (upper photosphere/lower
chromosphere) \citep{November1982,November1989}.
\cite{Giovanelli1980} found that the supergranular horizontal
velocities show no significant variation with height over the range
of formation of C~{\sc{i}} $\lambda$9111, Fe~{\sc{i}} $\lambda$8688,
and Mg~{\sc{i}} $\lambda$8806, but there is a substantial reduction
to about one-half of this at the level of Ca~{\sc{ii}}
$\lambda$8542. More recently, \cite{DelMoro2007} investigated the
3-D photospheric velocity field of a supergranular cell, and
proposed that large flow features can penetrate into the upper
photosphere.

\cite{Simon1964} suggested that the supergranulation flow field
carries the internetwork magnetic field to cell boundaries, forming
the pattern of the chromospheric network. So far many observations
have confirmed that most internetwork magnetic elements move toward
the boundaries of the magnetic network \citep[e.g.,][]{Martin1988,
Simon1988,WangZirin1988,Zhang1998}. Using magnetograms obtained by
the Big Bear Solar Observatory (BBSO), \cite{Zhang2006} estimated
the emergence rate of the new flux (ephemeral regions) to be
$\sim$5$\times10^{-19}$ cm$^{-2}$~day$^{-1}$ in the quiet Sun.
\cite{WangJ1995} calculated the mean area of internetwork magnetic
elements, which is about $10^{\prime\prime}$$^2$. Then the mean size
and height of an internetwork magnetic element should be about
$3^{\prime\prime}$. According to \cite{Vernazza1981}, the
temperature minimum, which marks the boundary between the
photosphere and chromosphere, is located at 500~km above
$\tau_{500}$=1. So in the internetwork regions, there should be a
large number of magnetic elements which extend above the
photosphere. And when these internetwork magnetic elements are swept
by the supergranular flows to the magnetic network, we should see
the horizontal motions of their footpoints and apices in the
photosphere and chromosphere, respectively. Unfortunately,
high-quality magnetograms of the quiet Sun above the photosphere are
still difficult to obtain. Thus, currently it is difficult to
validate the existence of these strong internetwork magnetic
elements above the photosphere. However, in an active region
\cite{Lin2006} found a correlation between localized emission
enhancement above the photosphere (in the chromosphere and TR) and a
moving magnetic feature, which followed the boundary of a
supergranulation cell. \cite{Zhang2000} analyzed quiet-Sun
photospheric and chromospheric magnetograms observed by the vector
video magnetograph at Huairou Solar Observing Station, and found
that all visible variations in the photosphere had corresponding
variations in the chromosphere, although the H$\beta$ line, which
they used to derive the chromospheric magnetogram, are complicated
and may include signals from both the photosphere and chromosphere.
\cite{Solanki2003} observed a set of rising magnetic loops and found
that magnetic signatures of the loop legs are clearly present in the
magnetogram of the upper chromosphere. Recently, based on Hinode
observations, \cite{Shimojo2009} even found a clear correlation
between coronal activities and photospheric minority magnetic
polarities in the polar coronal hole. All these observations seem to
suggest that horizontal movements of internetwork magnetic elements
should reveal horizontal motions both in and above the photosphere.

If the chromospheric internetwork grains have a one-to-one
correspondence to enhanced magnetic fields
\citep{Sivaraman1982,Dame1984,Dame1985,Dame1987,Dame1988,Sivaraman1991,Sivaraman2000},
or the persistent flashers which are related to strong internetwork
magnetic elements
\citep{Brandt1992,Brandt1994,Nindos1998,Krijger2001,McAteer2004,deWijn2005}
play a dominant role in the chromospheric internetwork emission, the
velocity fields revealed by TRACE UV observations naturally
represent the horizontal component of magneto-convection in the
chromosphere. By comparing the velocity fields obtained in white
light with those in the three UV passbands, we can conclude that the
supergranular pattern does not completely disappear above the
photosphere. The diverging flows in some cells are still very
clearly seen in the chromosphere. Some supergranular motions are
extending into the overlying formation layers of the 1700~{\AA},
1600~{\AA}, and 1550~{\AA} emission. The comparable velocities
derived in the photosphere and chromosphere are also easy to
understand because they reflect horizontal motions of the same
features, although at different heights.

The existence of a supergranule-scale magneto-convection in the
chromosphere may shed new light on the study of mass and energy
supply to the corona and solar wind at the height of the
chromosphere. It is believed that part of the magnetic network flux
opens into the corona in the shape of funnels, whilst the rest of
the network consists of a dense population of low-lying loops with
lengths less than $10^4$~km and varying orientations
\citep{Dowdy1986,Marsch1997,Peter2001}. Magnetic loops reaching to
the height of the chromosphere and TR can be swept by the
supergranular flow from the cell interior to its boundaries, where
they can interact with magnetic funnels, and by reconnection may
supply mass and energy to the funnels
\citep{Axford1999,Aiouaz2005,Aiouaz2008,He2007,McIntosh2007,Tian2008a,Tian2009}.
Reconnection at the interface between cool side loops and the
network flux tubes may occur in the chromosphere and TR, resulting
in the solar wind outflow
\citep{Tu2005,He2008,Buchner2005,Tian2008b,Tian2010} or upflows
along loop legs \citep{Tian2008a,Tian2009}, and downflows at lower
layers. Thus, horizontal motions representing supergranule-scale
magneto-convection in the chromosphere and TR are vital in such
processes as mass supply and energy delivery to the corona and solar
wind through funnels. Our finding may represent the observational
evidence for these motions.

However, we should point out that the extension of strong magnetic
elements into the chromosphere can only be explicitly confirmed when
the magnetic field above the photosphere can be accurately measured,
and that the existence of such supergranule-scale magneto-convection
in the chromosphere can only be confirmed by following the evolution
of the chromospheric magnetic field with high cadence and high
resolution.

\subsection{Chromospheric brightness evolution induced by photospheric granular motions}

As mentioned above, some authors doubt the magnetic origin of the
chromospheric cell grains, but believe that cell grains are caused
by weak acoustic shocks which propagate upwards through the low
chromosphere
\citep{Rutten1991,Remling1996,Carlsson1997,Loukitcheva2009}.
\cite{Lites1999} confirmed the existence of magnetism-related grains
but pointed out that they don't play a dominant role in the
chromospheric internetwork emission. If most cell grains are indeed
of acoustic origin, then the horizontal motions we derived by using
TRACE UV observations are likely to represent chromospheric
brightness evolution induced by the shocks.

In the one-dimensional simulation of \cite{Carlsson1997}, a piston
located at the bottom of the computational domain (100 km below
$\tau_{500}$=1) drives waves through the atmosphere, then
propagating waves near or just above the acoustic cutoff frequency
interfering with higher frequency waves that induce them to steepen
rapidly and form shocks near 1 Mm above $\tau_{500}$=1. The
chromospheric shocks produce a large source function, yielding the
high emissivity of internetwork grains. If this scenario is correct,
it is essential to investigate the horizontal distribution of
pistons at the bottom layer, which was not addressed in the
simulation of \cite{Carlsson1997} due to its one-dimensional
restriction.

Several attempts have been made to investigate this issue. As
pointed out by \cite{Rutten1999b}, turbulent convection seems to
supply the pistons that excite shock sequences and produce the cell
grains. \cite{Rimmele1995} found that enhanced acoustic wave
activities, the so-called "acoustic events", occur preferentially in
dark intergranular lanes. This spatial correspondence was later
confirmed by \cite{Goode1998} and \cite{Strous2000}.
\cite{Hoekzema1998} and \cite{HoekzemaEtal1998} further pointed out
that dark intergranular lanes tend to show excess $\sim$3-minute
waves in the photosphere. These transient acoustic waves are
suggested to be excited by small granules that undergo a rapid
collapse \citep{Rast1999,Skartlien2000}.

The relationship between enhanced wave activities in the photosphere
and enhanced chromospheric emission was also investigated by several
authors. For example, through wavelet analyses, \cite{Kamio2006}
found that the occurrence of chromospheric brightenings is
correlated with enhanced 5 mHz velocity oscillations in the
chromosphere and the photosphere. \cite{HoekzemaEtal1998} concluded
that the preferential alignment between $\sim$3-minute waves and
dark photospheric intergranular lanes does not survive to the
chromospheric heights; instead, there seems to be a correspondence
between excess chromospheric brightness and intergranular lanes at a
time delay of 2.5 minutes. \cite{Hoekzema2002} found that sites of
enhanced wave activity in the granulation preferentially co-locate
with exceptionally bright chromospheric internetwork grains, at a
delay of about 2 minutes which might represent the sound travel time
to the chromosphere. While through a statistic study,
\cite{Cadavid2003} found that 72\% of the G-band darkening events
are followed by an enhanced chromospheric emission 2 minutes later;
in the remaining 28\% cases, the timing is reversed.
\cite{Cadavid2003} also found that the G-band darkening events are
usually accompanied by transient enhancement of the measured
magnetic field, indicating collapse of intergranular lanes.
According to \cite{Goode2002}, a collapse in the intergranular lanes
can produce upward-propagating waves which subsequently lead to
chromospheric brightenings.

The horizontal motions we derived in the chromosphere show a similar
pattern to the photospheric supergranulation. If most chromospheric
cell grains are produced by acoustic shocks, our results should
indicate that the acoustic shocks are modulated by photospheric or
subsurface flows. \cite{Wedemeyer2004} pointed out that convection
motions play an important role in the excitation of acoustic waves.
It is well known that photospheric granules tend to move in a
systematic way as characterized by the supergranulation. If the
collapse of intergranular lanes is really the source of acoustic
waves that produce the chromospheric cell grains, we may expect a
coherence between the photospheric supergranulation and the motion
of the collapsing site of intergranular lanes. In other words, the
collapsing site of intergranular lanes has a tendency to move with
the granules. This will also naturally yield a comparable horizontal
velocity in the photosphere and chromosphere.

As pointed out by \cite{Hoekzema1998}, it is worthwhile to use
photospheric flow tracking to enable studies of chromospheric
response to photospheric or subsurface excitation sites while
following migrations of the latter over mesoscale and larger
distances. Our study focuses on the comparison between horizontal
motions in the photosphere and chromosphere, and thus provide some
insights into this issue. The coherent supergranule-scale behavior
between motions in the two layers seems to indicate that the
chromospheric cell grains mark locations where acoustic events
follow on granular collapses in the evolving intergranular lanes.

It is also possible that the propagation of the shock waves is
influenced by the supergranulation. The shock wave flux might be
advected by the supergranular flows toward the network while it is
propagating upward from the photosphere and dissipated in the
chromosphere.

However, the resolution of the TRACE data we used here is not very
high. More high-resolution and high-cadence observations of
different layers are needed to further study the dynamics of
granular evolution in intergranular lanes and the chromospheric
response \citep{Rutten2008}. And detailed 3-dimensional numerical
simulation should also be done to investigate the role of granular
motions in the excitation of acoustic shocks and the subsequent
production of chromospheric cell grains.

\section{Summary and conclusion}

We have applied the highly efficient balltracking technique to TRACE
images obtained in the white-light band and three UV passbands
centered at 1700~\AA, 1600~\AA, and 1550~\AA. We have demonstrated
that the supergranular flow pattern in the photosphere can be
recovered by applying this tracking method to the white-light images
observed by TRACE. This is the first time that horizontal motions in
a large quiet area of the solar chromosphere have been investigated.
Our analysis revealed a striking correlation between the horizontal
velocities derived in the white-light band and the UV passbands.

The interpretation of our finding is not straightforward, since we
tracked the apparent motions of the chromospheric internetwork
(cell) grains, the nature of which is still under debate. If the
cell grains (or most of them) correspond to enhanced internetwork
magnetic elements, the velocity fields revealed by TRACE UV
observations should represent the horizontal component of
magneto-convection in the chromosphere. Then our finding seems to
provide evidence for the way on which mass and energy are supplied
to the corona and solar wind at the height of the chromosphere,
which is predicted or suggested by many recent observational and
modeling studies. However, as believed by many authors, the cell
grains may be entirely produced by acoustic shocks propagating
upward to the chromosphere. If the cell grains are indeed of
acoustic origin, the velocity fields revealed by our TRACE UV
observations should reflect the motion pattern of the short-living
chromospheric brightness as induced by acoustic shocks. Then the
striking correlation between the horizontal velocities derived in
the UV and white-light passbands seems to indicate that the
excitation of enhanced wave activity is modulated by the
photospheric granular motions, or that the shock waves are advected
by the supergranular flows toward the network while it is
propagating upward from the photosphere and dissipated in the
chromosphere.

We conclude that it is important to investigate the role of granular
motions in the excitation of shocks through numerical modeling. In
addition, future high-resolution and high-cadence observations
including dopplergrams, magnetograms, and imaging of the photosphere
and chromosphere are needed to investigate and understand the
coupling between the two layers.

We realize that our results and conclusions need to be checked and
confirmed by future observations and with improved techniques. The
mission of the Solar Dynamic Observatory (SDO), which was launched
in February 2010, will provide full-disk high-resolution
photospheric magnetograms and chromospheric images. It is likely
that these data will better our understanding of the evolution of
the chromospheric emission.

\begin{acknowledgements}
The TRACE satellite is a NASA small explorer mission that images the
solar photosphere, transition region and corona with unprecedented
spatial resolution and temporal continuity. Hui Tian and Raphael
Attie are supported by the IMPRS graduate school run jointly by the
Max Planck Society and the Universities of G\"ottingen and
Braunschweig. The work of Hui Tian's group at Peking University is
supported by the National Natural Science Foundation of China (NSFC)
under contracts 40874090, 40931055, and 40890162. The space physics
group at PKU are also supported by the Beijing Education Project
XK100010404, the Fundamental Research Funds for the Central
Universities, and the National Basic Research Program of China under
grant G2006CB806305. We thank the anonymous referee for his/her
careful reading of the paper and for the comments and suggestions.
\end{acknowledgements}


\begin{thebibliography}{}
\bibitem[Aiouaz et al.(2005)]{Aiouaz2005}
Aiouaz, T., Peter, H., \& Lemaire P. 2005, A\&A, 435, 713
%
\bibitem[Aiouaz(2008)]{Aiouaz2008}
Aiouaz, T., 2008, ApJ, 674, 1144
%
\bibitem[Axford et al.(1999)]{Axford1999}
Axford, W. I., Mckenzie, J. F., \& Sukhorukova, G. V. 1999, Space Science Reviews, 87, 25
%
\bibitem[Attie et al.(2009)]{Attie2009}
Attie, R., Innes, D. E., \& Potts, H. E. 2009, A\&A, 493, L13
%
\bibitem[Berrilli et al.(2002)]{Berrilli2002}
Berrilli, F., Consolini, G., Pietropaolo, E., et al. 2002, A\&A,
381, 253
%
\bibitem[Brandt et al.(1992)]{Brandt1992}
Brandt, P. N., Rutten, R. J., Shine, R. A., \& Trujillo Bueno, J.
1992, in Cool Stars, Stellar Systems, and the Sun, ed. M. S.
Giampapa, \& J. A. Bookbinder, ASP Conf. Ser., 26, 161
%
\bibitem[Brandt et al.(1994)]{Brandt1994}
Brandt, P. N., Rutten, R. J., Shine, R. A., \& Trujillo Bueno, J.
1994, in Solar Surface Magnetism, ed. R. J. Rutten, \& C. J.
Schrijver, NATO ASI Series C 433, Kluwer, Dordrecht, 251
%
\bibitem[B\"{u}chner \& Nikutowski(2005)]{Buchner2005}
B\"{u}chner, J. \& Nikutowski, B. 2005, Proceedings of Solar Wind 11/SOHO 16, ESA SP-592, 141
%
\bibitem[Cadavid et al.(2003)]{Cadavid2003}
Cadavid, A. C., Lawrence, J. K., Berger, T. E., \& Ruzmaikin, A.
2003, ApJ, 586, 1409
%
\bibitem[Carlsson \& Stein(1997)]{Carlsson1997}
Carlsson, M., \& Stein, R. F. 1997, ApJ, 481, 500
%
\bibitem[Chae et al.(2000)]{Chae2000}
Chae, J., Denker, C., Spirock, T. J., et al. 2000, Sol. Phys., 195, 333
%
\bibitem[Curdt et al.(2008)]{Curdt2008}
Curdt, W., Tian, H., Dwivedi, B. N., \& Marsch, E. 2008, A\&A, 491,
L13
%
\bibitem[Dammasch et al.(2008)]{Dammasch2008}
Dammasch, I. E., Curdt, W., Dwivedi, B.~N., \& Parenti, S. 2008,
Ann. Geophys., 26, 2955
%
\bibitem[Dam\'{e} et al.(1984)]{Dame1984}
Dam\'{e}, L., Gouttebroze, P., \& Malherbe, J.-M. 1984, A\&A, 130,
331
%
\bibitem[Dam\'{e}(1985)]{Dame1985}
Dam\'{e}, L. 1985, in Theoretical Problems in High Resolution Solar
Physics, ed. H. U. Schmidt, MPA/LPARL Workshop (M\"{u}nchen:
Max-Planck-Institut f\"{u}r Physik und Astrophysik MPA 212), 244
%
\bibitem[Dam\'{e} \& Martic(1987)]{Dame1987}
Dam\'{e}, L., \& Martic, M. 1987, ApJ, 314, L15
%
\bibitem[Dam\'{e} \& Martic(1988)]{Dame1988}
Dam\'{e}, L., \& Martic, M. 1988, in Advances in Helio- and
Asteroseismology, ed. J. Christensen-Dalsgaard, \& S. Frandsen
(Dordrecht: Reidel), IAU Symp., 123, 433
%
\bibitem[Del Moro(2007)]{DelMoro2007}
Del Moro, D., Giordano, S., \& Berrilli, F. 2007, 472, 599
%
\bibitem[Deubner(1971)]{Deubner1971}
Deubner, F.-L. 1971, Sol. Phys., 17, 6
%
\bibitem[de Wijn et al.(2005)]{deWijn2005}
de Wijn, A. G., Rutten, R. J., Haverkamp, E. M. W. P., \&
S\"{u}tterlin, P. 2005, A\&A, 441, 1183
%
\bibitem[de Wijn et al.(2008)]{deWijn2008}
de Wijn, A. G., et al. 2008, ApJ, 684, 1469
%
\bibitem[Dowdy et al.(1986)]{Dowdy1986}
Dowdy, J. F. Jr., Rabin, D., \& Moore, R. L. 1986, Sol. Phys., 105,
35
%
\bibitem[Foukal(1978)]{Foukal1978}
Foukal, P., 1978, ApJ, 223, 1046
%
\bibitem[Gabriel(1976)]{Gabriel1976}
Gabriel, A. H. 1976, Philos. Trans. R. Soc. London A, 281, 575
%
\bibitem[Giovanelli(1980)]{Giovanelli1980}
Giovanelli, R. G., 1980, Sol. Phys., 67, 211
%
\bibitem[Goode et al.(1998)]{Goode1998}
Goode, P., Strous, L., Rimmele, T., \& Stebbins, R. 1998, ApJ, 495,
L27
%
\bibitem[Goode (2002)]{Goode2002}
Goode, P. R. 2002, Bulletin of the American Astronomical Society,
34, 730
%
\bibitem[Handy et al.(1999)]{Handy1999}
Handy, B. N., Acton, L. W., Kankelborg, C. C., et al. 1999, Sol. Phys., 187, 229
%
\bibitem[He et al.(2007)]{He2007}
He, J.-S., Tu, C.-Y., \& Marsch, E. 2007, A\&A, 468, 307
%
\bibitem[He et al.(2008)]{He2008}
He, J.-S., Tu, C.-Y., \& Marsch, E. 2008, Sol. Phys., 250, 147
%
\bibitem[Hoekzema \& Rutten(1998)]{Hoekzema1998}
Hoekzema, N. M., \& Rutten, R. J. 1998, A\&A, 329, 725
%
\bibitem[Hoekzema et al.(1998)]{HoekzemaEtal1998}
Hoekzema, N. M., Rutten, R. J., Brandt, P. N., \& Shine, R. A. 1998,
A\&A, 329, 276
%
\bibitem[Hoekzema et al.(2002)]{Hoekzema2002}
Hoekzema, N. M., Rimmele, T. R., \& Rutten, R. J. 2002, A\&A, 390,
681
%
\bibitem[Innes et al.(2009)]{Innes2009}
Innes, D. E., Genetelli, A., Attie, R., \& Potts, H. E. 2009, A\&A, 495, 319
%
\bibitem[Jin et al.(2009)]{Jin2009}
Jin, C.-L., Wang, J.-X., \& Zhao, M. 2009, ApJ, 690, 279
%
\bibitem[Kamio et al.(2006)]{Kamio2006}
Kamio, S., \& Kurokawa, H. 2006, A\&A, 450, 351
%
\bibitem[Krijger et al.(2001)]{Krijger2001}
Krijger, J. M., Rutten, R. J., Lites, B. W., et al. 2001, A\&A, 379,
1052
%
\bibitem[Krijger et al.(2002)]{Krijger2002}
Krijger, J. M., Roudier, T., \& Rieutord, M. 2002, A\&A, 387, 672
%
\bibitem[Leighton et al.(1962)]{Leighton1962}
Leighton, R. B., Noyes, R. W., \& Simon, G. W. 1962, ApJ, 135, 474
%
\bibitem[Lin et al.(2006)]{Lin2006}
Lin, C.-H., Banerjee, D., O'Shea, E., \& Doyle, J. G. 2006, A\&A,
460, 597
%
\bibitem[Lites et al.(1999)]{Lites1999}
Lites, B. W., Rutten, R. J., \& Berger, T. E. 1999, ApJ, 517, 1013
%
\bibitem[Livingston \& Harvey(1975)]{Livingston1975}
Livingston, W. C., \& Harvey, J. 1975, AAS Bull. 7, 346
%
\bibitem[Loukitcheva et al.(2009)]{Loukitcheva2009}
Loukitcheva, M., Solanki, S. K., \& White, S. M. 2009, A\&A, 497,
273
%
\bibitem[Marsch \& Tu(1997)]{Marsch1997}
Marsch, E., \& Tu, C.-Y. 1997, Sol. Phys., 176, 87
%
\bibitem[Marsch et al.(2008)]{Marsch2008}
Marsch, E., Tian, H., Sun, J., Curdt, W., \& Wiegelmann, T. 2008,
ApJ, 684, 1262
%
\bibitem[Martin(1988)]{Martin1988}
Martin, S. F. 1988, Sol. Phys., 117, 243
%
\bibitem[McAteer et al.(2004)]{McAteer2004}
McAteer, R. T. J., Gallagher, P. T., Bloomfield, D. S., et al. 2004, ApJ, 602, 436
%
\bibitem[McIntosh et al.(2007)]{McIntosh2007}
McIntosh, S. W., Davey, A. R., Hassler, D. M., et al. 2007, ApJ,
654, 650
%
\bibitem[Nindos \& Zirin(1998)]{Nindos1998}
Nindos, A., \& Zirin, H. 1998, Sol. Phys., 179, 253
%
\bibitem[November et al.(1979)]{November1979}
November, L. J., Toomre, J., \& Gebbie, K. B. 1979, ApJ, 227, 600
%
\bibitem[November et al.(1982)]{November1982}
November, L. J., Toomre, J., Gebbie, K. B., \& Simon, G. W. 1982,
ApJ, 258, 846
%
\bibitem[November \& Simon(1988)]{November1988}
November, L. J., \& Simon, G. W. 1988, ApJ, 333, 427
%
\bibitem[November(1989)]{November1989}
November, L. J. 1989, ApJ, 344, 494
%
\bibitem[Peter(2001)]{Peter2001}
Peter, H. 2001, A\&A, 374,1108
%
\bibitem[Potts et al.(2004)]{Potts2004}
Potts, H. E., Barrett, R. K., \& Diver, D. A. 2004, A\&A, 424, 253
%
\bibitem[Potts et al.(2007)]{Potts2007}
Potts, H. E., Khan, J. I., \& Diver, D. A. 2007, Sol. Phys., 245, 55
%
\bibitem[Potts \& Diver(2008)]{Potts2008}
Potts, H. E., \& Diver, D. A. 2008, Sol. Phys., 248, 263
%
\bibitem[Rast(1999)]{Rast1999}
Rast, M. P. 1999, ApJ, 524, 462
%
\bibitem[Remling et al.(1996)]{Remling1996}
Remling, B., Deubner, F.-L., \& Steffens, S. 1996, A\&A, 316, 196
%
\bibitem[Rieutord et al.(2001)]{Rieutord2001}
Rieutord, M., Roudier, T., Ludwig, H.-G., et al. 2001, A\&A, 377, L14
%
\bibitem[Rimmele et al.(1995)]{Rimmele1995}
Rimmele, T. R., Goode, P. R., Harold, E., \& Stebbins, R. T. 1995,
ApJ, 444, L119
%
\bibitem[Roudier et al.(1999)]{Roudier1999}
Roudier, T., Rieutord, M., Malherbe, J. M., \& Vigneau, J. 1999,
A\&A, 349, 301
%
\bibitem[Rutten \& Uitenbroek(1991)]{Rutten1991}
Rutten, R. J., \& Uitenbroek, H. 1991, Sol. Phys., 134, 15
%
\bibitem[Rutten et al.(1999a)]{Rutten1999a}
Rutten, R. J., de Pontieu, B., \& Lites, B. W. 1999a, in High
Resolution Solar Physics: Theory, Observations, and Techniques, ed.
T. R. Rimmele, K. S. Balasubramaniam, \& R. R. Radick, Procs. 19th
NSO/Sacramento Peak Summer Workshop, ASP Conf. Ser., 183, 383
%
\bibitem[Rutten et al.(1999b)]{Rutten1999b}
Rutten, R. J., Lites, B. W., Berger, T. E., \& Shine, R. A. 1999b,
in Solar and stellar activity: Similarities and Differences, ed. C.
J. Butler, \& J. G. Doyle, ASP Conf. Ser., 158, 249
%
\bibitem[Rutten \& Krijger(2003)]{Rutten2003}
Rutten, R. J., \& Krijger, J. M. 2003, A\&A, 407, 735
%
\bibitem[Rutten et al.(2004)]{Rutten2004}
Rutten, R. J., de Wijn, A. G., \& S\"{u}tterlin, P. 2004, A\&A, 416,
333
%
\bibitem[Rutten et al.(2008)]{Rutten2008}
Rutten, R. J., van Veelen, B., \& S\"{u}tterlin, P. 2008, Sol.
Phys., 251, 533
%
\bibitem[Shimojo \& Tsuneta(2009)]{Shimojo2009}
Shimojo, M, \& Tsuneta, S. 2009, ApJ, 706, L145

\bibitem[Simon \& Leighton(1964)]{Simon1964}
Simon, G. W., \& Leighton, R. B. 1964, ApJ, 140, 1120
%
\bibitem[Simon et al.(1988)]{Simon1988}
Simon, G. W., et al. 1988, ApJ, 327, 964
%
\bibitem[Sivaraman \& Livingston(1982)]{Sivaraman1982}
Sivaraman, K. R., \& Livingston, W. C. 1982, Sol. Phys., 80, 227
%
\bibitem[Sivaraman(1991)]{Sivaraman1991}
Sivaraman, K. R. 1991, in Mechanisms of Chromospheric and Coronal
Heating, ed. P. Ulmschneider, E. Priest, \& B. Rosner, Heidelberg
Conf. (Berlin: Springer Verlag), 44
%
\bibitem[Sivaraman et al.(2000)]{Sivaraman2000}
Sivaraman, K. R., Gupta, S. S., Livingston, W. C., et al. 2000,
A\&A, 363, 279
%
\bibitem[Skartlien et al.(2000)]{Skartlien2000}
Skartlien, R., Stein, R. F., \& Nordlund, {\AA} 2000, ApJ, 541, 468
%
\bibitem[Solanki et al.(2003)]{Solanki2003}
Solanki, S. K., Lagg, A., Woch, J., Krupp, N., \& Collados, M. 2004,
Nature, 425, 692
%
\bibitem[Strous et al.(2000)]{Strous2000}
Strous, L. H., Goode, P. R., \& Rimmele, T. R. 2000, ApJ, 535, 1000
%
\bibitem[Tian et al.(2008a)]{Tian2008a}
Tian, H., Tu, C.-Y., Marsch, E., He, J.-S., \& Zhou, G.-Q. 2008a,
A\&A, 478, 915
%
\bibitem[Tian et al.(2008b)]{Tian2008b}
Tian, H., Xia, L.-D., He, J.-S., Tan, B., Yao, S. 2008b, Chin. J.
Astron. Astrophys., 8, 732
%
\bibitem[Tian et al.(2009)]{Tian2009}
Tian, H., Marsch, E., Curdt, W., He, J.-S. 2009, ApJ, 704, 883
%
\bibitem[Tian et al.(2010)]{Tian2010}
Tian, H., Tu, C.-Y., Marsch, E., He, J.-S., \& Kamio, S. 2010, ApJ,
709, L88
%
\bibitem[Title et al.(1989)]{Title1989}
Title, A. M., Tarbell, T. D., Topka, K. P., et al. 1989, ApJ, 336,
475
%
\bibitem[Tritschler et al.(2007)]{Tritschler2007}
Tritschler, A., Schmidt, W., Uitenbroek, H., \& Wedemeyer-B\"{o}hm,
S. 2007, A\&A, 462, 303
%
\bibitem[Tu et al.(2005)]{Tu2005}
Tu, C.-Y., Zhou, C., Marsch, E., Xia, L.-D., Zhao, L., Wang, J.-X.,
Wilhelm, K. 2005, Science, 308, 519
%
\bibitem[Vernazza et al.(1981)]{Vernazza1981}
Vernazza, J. E., Avrett, E. H., \& Loeser, R. 1981, ApJS, 45, 635
%
\bibitem[Wang \& Zirin(1988)]{WangZirin1988}
Wang, H., \& Zirin, H. 1988, Sol. Phys., 115, 205
%
\bibitem[Wang(1988)]{Wang1988}
Wang, H. 1988, Sol. Phys., 117, 343
%
\bibitem[Wang et al.(1995)]{WangJ1995}
Wang, J., Wang, H., Tang, F., Lee, J. W., \& Zirin, H. 1995, ApJ,
160, 277
%
\bibitem[Wang et al.(1995)]{Wang1995}
Wang, Y., Noyes, R. W., Tarbell, T. D., \& Title, A. M. 1995, ApJ,
447, 419
%
\bibitem[Wedemeyer et al.(2004)]{Wedemeyer2004}
Wedemeyer, S., Freytag, B., Steffen, M., Ludwig, H.-G., \& Holweger,
H. 1999, A\&A, 414, 1121
%
\bibitem[Worden et al.(1999)]{Worden1999}
Worden, J., Harvey, J., \& Shine, R. 1999, ApJ, 523, 450
%
\bibitem[Yang et al.(2003)]{Yang2003}
Yang, G., Xu, Y., Wang, H., \& Denker, C. 2003, ApJ, 597, 1190
%
\bibitem[Yang et al.(2009)]{Yang2009}
Yang, S. H., Zhang, J., Jin, C. L., Li, L. P., \& Duan, H. Y. 2009,
A\&A, 501, 745
%
\bibitem[Yi \& Molowny-Horas(1995)]{Yi1995}
Yi, Z., \& Molowny-Horas, R. 1995, A\&A, 295, 199
%
\bibitem[Zhang et al.(1998)]{Zhang1998}
Zhang, J., Wang, J., Wang, H., \& Zirin, H. 1998, A\&A, 335, 341
%
\bibitem[Zhang et al.(2006)]{Zhang2006}
Zhang, J., Ma, J., \& Wang, H. 2006, ApJ, 649, 464
%
\bibitem[Zhang \& Zhang(2000)]{Zhang2000}
Zhang, M., \& Zhang, H. 2000, Sol. Phys., 194, 19
%
\end{thebibliography}
\end{document}